\documentclass[aps,prd,reprint,nofootinbib,amsmath,amssymb]{revtex4-2}

\usepackage{graphicx,amsmath,amssymb,amstext}
\usepackage{amssymb,amsbsy,amsfonts,amsthm,hyperref}
\usepackage{url}
\usepackage[normalem]{ulem}
\usepackage{subcaption}

\usepackage{placeins}

\newcommand{\adsurl}[1]{\href{#1}{ADS}}
\providecommand{\url}[1]{\href{#1}{#1}}

\newcommand{\be}{\begin{equation}}
\newcommand{\ee}{\end{equation}}
\newcommand{\bea}{\begin{eqnarray}}
\newcommand{\eea}{\end{eqnarray}}

\usepackage{color}
\usepackage{ifthen}
\newboolean{editorial}
\setboolean{editorial}{true}
\newcommand{\editorial}[2]{\ifthenelse{\boolean{editorial}}{\textcolor{red}{[\textsf{\textbf{{#1}}}: }\textcolor{blue}{\textsf{{#2}}}\textcolor{red}{]}}{}}

\usepackage{ulem}

\definecolor{amber}{rgb}{1.0, 0.49, 0.0}

\begin{document}

\title{No persuasive evidence yet of gravitational-wave tails from perturbers along the line of sight in LVK observations}

\author{Kuba Kopczuk${}^{1}$}
\author{Ethan Blake${}^{1}$}
\author{Matthew F. Carney${}^{2}$}
\author{Jeremy G. Baier${}^{3}$}
\author{Michael Van Keuren${}^{1}$}
\author{Adam Khan${}^{1}$}
\author{Leslie Wade${}^{1}$}
\author{Madeline Wade${}^{1}$}
\author{Glenn Starkman${}^{4}$}
\author{Craig Copi${}^{4}$}

\affiliation{${}^1$Department of Physics, Kenyon College, 201 N College Rd, Gambier, OH 43022, USA}
\affiliation{${}^2$Department of Physics and McDonnell Center for the Space Sciences,
Washington University, St. Louis, MO 63130, USA}
\affiliation{${}^3$Department of Physics, Oregon State University, Corvallis, OR 97331, USA}
\affiliation{${}^4$Department of Physics/CERCA/Institute for the Science of Origins, Case Western Reserve University, Cleveland, OH 44106-7079 -- USA}

\date{\today}

\begin{abstract}
Observations of compact binary coalescences (CBCs) by the  LIGO-Virgo-KAGRA collaboration resulted in 90 events in the first three observing runs. 
We search the data near the loudest of these detections for so-called ``gravitational glints,'' by performing Bayesian model comparison. 
Gravitational glints are gravitational-wave tails caused by interactions between a signal and a spacetime perturber and are theorized to present as echoes of the primary signal.
We do not find convincing evidence of gravitational glints in binary black hole or intermediate-mass black hole binary events with a signal-to-noise ratio (SNR) of at least 12.
We also found that standard matched-filter CBC searches with template banks that do not contain gravitational glints would likely not have missed signals containing a gravitational glint.
We therefore estimate upper limits on the probability of a glint as a function of its relative amplitude and use these estimates to constrain the properties of perturbers in the Universe. 
With increased sensitivity in the fourth observing run, which will increase the number and SNR of detectable events, we remain optimistic that glints will soon be detectable.

\end{abstract}

\maketitle

\section{Introduction}

In addition to the detection and subsequent inference of source parameters of astrophysical systems, the measurements of gravitational waves (GW) achieved by the LIGO-Virgo-Kagra (LVK) collaboration open the door to probing phenomena within General Relativity (GR) that have not yet been observed.
One such GR effect is the  prediction of detectable GW ``tails'' \cite{Copi:2022ire} generated by  interactions between the primary GW signal and the gravitational potential of a massive perturber of the background flat (or nearly flat) Lema\^{i}tre-Friedmann-Robertson-Walker geometry, such as a star or compact stellar remnant.  
Meanwhile, one can hope as well for effects outside GR that may or may not have been anticipated. 

Radiation tails have been, for over a hundred years, a known mathematical feature of the Green's functions of massless fields propagating in geometries that are not conformally flat \cite{hadamard1923lectures}.
The Green's function contains two contributions---the usual delta function of the light cone, which describes speed-of-light propagation, and a theta function of the light cone, which describes slower-than-c propagation and is known as the radiation tail. 
In the context of small perturbations in the geometry around a homogeneous conformally flat background, the radiation tail can be understood as the coherent scattering of the propagating wave off the perturbations.

The tail, or the theta-function part of the Green's function, is an established feature of the standard gravitational waveform.
The backscattering of an inspiraling binary's radiation off the curvature associated with the system's own mass produces a hereditary, non-local-in-time contribution that enters the phasing at 1.5PN order and corresponds to a tail contribution to the radiation-reaction force \cite{Blanchet1992br}. 
This contribution has been confirmed indirectly through GW observations: the 1.5PN deviation parameter, where the dominant tail effect enters, is now constrained at the percent level by GW250114 \cite{GW250114_TGR} and across the GWTC-4.0 catalog \cite{LIGOScientific:2026fcf}.
What has not been observed is a resolved, localized manifestation of the same effect: a discrete tail produced when radiation backscatters off the curvature of an individual compact object lying close to the line of sight, rather than off the cumulative field of the emitting source. 
In the electromagnetic case such localized tails are extremely weak, because the relevant small parameter is the ratio of the radiation's wavelength to the curvature scale of the geometry it traverses. 
Electromagnetic radiation with wavelength longer than $\sim 100$~m does not penetrate the solar system due to the ionized solar wind, while curvature scales that small are rare except inside electromagnetically opaque compact objects.
Electromagnetic radiation tails are therefore extremely weak.

The detectability of GWs \cite{LIGOScientific:2016aoc} potentially changes this situation.  
GWs are unimpeded by the solar wind, and are able to probe the geometry inside a compact source, such as a star or any  stellar remnant other than a black hole.
Ref.~\cite{Copi:2022ire} demonstrated that at the GW frequencies probed by the LVK interferometers, such perturbers lying close to the lines-of-sight between us and distant GW sources should occasionally result in GW tails, known as  GW ``glints", that are faithful echoes of primary (i.e., light-cone) signals, with amplitudes only modestly reduced from the primary signal, delayed by a fraction of a wave period.

The detectability of glints raises a natural question about their relationship to gravitational lensing, which similarly produces time-delayed copies of a primary GW signal due to the influence of a massive perturber. 
Dedicated LVK lensing searches target repeated images and wave-optics signatures of isolated point-mass lenses \cite{Hannuksela:2019kle, LIGOScientific:2021izm, Abbott_2024}, and are sensitive to perturbers lying within a few Einstein radii of the line of sight. 
Closer analogs are population-microlensing and millilensing scenarios, in which smaller-scale structure along the line of sight modulates the waveform: stellar-mass microlenses embedded in the deflection field of a galaxy or cluster macromodel near a critical curve produce millisecond-scale time delays between microimages \cite{Diego:2019rzc}, while millilensing analyses model a small-scale lens that splits the signal into several overlapping copies whose interference imprints a frequency-dependent beating pattern \cite{Liu2023millilensing}. 
Indeed, the glint waveform is formally a special case of the phenomenological millilensing parameterization of Ref.~\cite{Liu2023millilensing}, with two component signals, relative magnification $\varepsilon^2$, and vanishing Morse phases. 
The underlying physics is nevertheless distinct. 
These lensing signals are all built from the on-light-cone (delta-function) piece of the Green's function, with the macromodel (where present) supplying the geometric leverage that brings stellar-mass lenses into the observable regime, and milli-images generically acquire nontrivial Morse phases that can distort the waveform. 
The glint is instead built from the theta-function piece---the radiation tail---with its observability set by the geometric scaling of the tail contribution from a single extended perturber, and it occurs for perturbers \emph{outside} their Einstein radius, where lensing produces no detectable additional images and these scenarios do not apply.

Glints also resemble the GW ``echoes'' predicted for horizonless exotic compact objects, in which reflections off near-horizon structure produce repeated, damped copies of the signal---a signature of physics beyond GR \cite{Abedi:2016hgu}. 
The LVK has searched for echoes in each catalog release, most recently in the GWTC-4.0 tests of general relativity \cite{LVK_TGR_III}, using both the phenomenological Abedi--Dykaar--Afshordi (ADA) template family \cite{Abedi:2016hgu} and templates motivated by black-hole perturbation theory (BHP) \cite{Testa:2018bzd}. 
A glint is morphologically close to an ADA echo train trimmed to a single echo ($N=1$) with amplitude $\varepsilon$, but the templates differ in detail: ADA echoes invert their phase at each reflection, repeat only the merger--ringdown portion of the waveform (truncated at a time set by an additional free parameter), and form a train of copies damped by a decay factor at each reflection, whereas the tail physics of Sec.~\ref{sec:tails} predicts a single, phase-preserving copy of the full inspiral--merger--ringdown signal.
The BHP templates differ further: their frequency-dependent reflectivity distorts each echo, in contrast to the frequency-independent relative amplitude of a glint, so BHP non-detections do not directly constrain glints. 
The interpretations are also entirely distinct---echoes probe the near-horizon structure of the merger remnant, while glints are a pure-GR propagation effect sourced by a perturber along the line of sight.

In the corner of parameter space where the ADA and glint templates most nearly coincide, echo non-detections are broadly complementary to the constraints we place here. 
Nevertheless, glints populate a region of parameter space that no existing LVK lensing or echo analysis targets: none applies priors appropriate to glints, and none interprets its results in terms of the line-of-sight perturber populations that could produce them---which is the aim of the present work.

For this study, we investigated all publicly released GW events from the first three LVK observing runs (O1-O3) in the third Gravitational Wave Transient Catalog (GWTC-3) \cite{GWTC-3} with signal-to-noise ratios (SNR) of at least 12, which included 33 binary black hole (BBH) events and one intermediate mass black hole (IMBH) event, for evidence of GW glints.
Although the explored LVK data set does not include a large enough sample of high SNR sources to expect a GW glint detection, we report here on a first attempt to characterize the ability to detect GW glints. Additionally, we studied the impact of GW glints on the sensitivity of modeled searches for compact binary mergers and found that for most combinations of GW glint parameters, the sensitivity of modeled searches should not be significantly impacted by the presence of GW glints.  We therefore concluded that an independent search over LVK data for compact merger events including GW glints in the template waveforms was not necessary.

Having examined 34 mergers with SNRs between 12.0 and 26.8 we report that none have statistically significant evidence of GW glints.
Even for the loudest events in our sample, a glint must have a relative amplitude of at least $\sim$22\% of the primary signal to be detectable.
Using this result, we are able to place constraints on the perturber population that could produce GW glints.

In Sec.~\ref{sec:tails} we review the physical motivation for GW glints.  In Sec.~\ref{sec:pe} we discuss the methods and results for determining evidence for a GW glint among confirmed GW compact binary merger events found in GWTC-3 \cite{GWTC-3}.  In Sec.~\ref{sec:search} we show the impact of GW glints on the sensitivity of modeled searches for compact binary merger objects.
In Sec.~\ref{sec:pe_ul} we discuss the upper limits on GW glint parameters and perturber density given the population of explored signals, and finally we conclude in Sec.~\ref{sec:discussion}.

\section{Review of Gravitational-Wave Tail Physics}
\label{sec:tails}

The determination that light propagates in vacuum at a single speed, $c$, regardless of the velocity of any inertial (constant velocity) observer of that light \cite{Michelson:1887zz} is one of the key paradigm-shifting experimental results underlying modern physics. 
Einstein, soon after his proposal of GR, recognized that it predicted that GWs should exist and also travel in empty space at this same unique velocity for all inertial observers  \cite{1916SPAW.......688E} . 
This equality of speeds of gravitational and electromagnetic waves (EMW) has recently been tested \cite{LIGOScientific:2017zic} to nearly $(1-3)\times10^{-15}$ by leveraging the almost-simultaneous arrival of gravitational and electromagnetic signals from a merging pair of neutron stars \cite{LIGOScientific:2017vwq,LIGOScientific:2017zic}. 

It is not surprising then that there is a widespread misconception that massless particles travel at the speed of light in vacuum, or equivalently that the vacuum Green's function of the electromagnetic potential $A_\mu$ \cite{Copi:2020qur} (or of the electromagnetic field strength tensor $F_{\mu\nu}$) and of transverse-traceless metric perturbations $h_{\mu\nu}^{TT}$  \cite{Copi:2022ire} (or of the Riemann tensor $R_{\mu\nu\alpha\beta}$) are proportional to a delta-function of the future light cone of the source.
Integrating such a Green's function against an electromagnetic current or a stress-energy current would consequently yield an information-, energy- and momentum-carrying disturbance only on the future light cones of the support of the source currents.

Most physicists, not to mention the public, are unaware that this is true only for EMW and GW propagating in even-dimensional (e.g. $3+1$-d)  Minkowski spacetime or other conformally flat even-dimensional spacetimes (such as unperturbed flat Lema\^{i}tre-Friedmann-Robertson-Walker spacetime).  
Otherwise, the Green's functions include two contributions---the well-known delta function and an additional contribution that is proportional to a theta function of the light-cone.  
Integrating these against a source yields an in-principle observable disturbance in the electromagnetic field or metric  for all times on or inside the future light cone of the support of the disturbance.  

To the best of our knowledge, the mathematical description of this ``tail'' phenomenon was first put forward by Hadamard \cite{hadamard1923lectures} just over a century ago. 
Much later, DeWitt and Brehme \cite{DewittBrehme1960} noted that ``a sharp pulse of light, when propagating in a curved 4-dimensional hyperbolic Riemannian manifold, does not, in general, remain a sharp pulse, but gradually develops a `tail'.''
Thorne and Kovacs \cite{thorne} developed a Green's function approach to calculate the observed radiation from a source in a weakly curved spacetime,  incorporating what they called ``tail radiation.''
This approach has been further developed by several authors  \cite{Nolan:1997jy,Pfenning_2002,Poisson2011,Chu:2011aa,Harte:2013dba,Chu:2020aa},
while alternatives have also been proposed \cite{Garoffolo:2022usx,Braga:2024pik}.

These formalisms have been applied to study the resulting self-force, which  DeWitt and DeWitt \cite{DeWitt:1964aa} first recognized in the context of the electromagnetic force in curved spacetime and described as arising ``from a back-scatter process in which the Coulomb field of the particle, as it sweeps over the `bumps' in space-time, receives `jolts' which are propagated back to the particle.''
(See also \cite{MinoSasakiTanaka1997,QuinnWald1997}.)
Meanwhile, others have studied the potential direct detection of the tail radiation for EMWs or GWs as they propagate through an inhomogeneous universe \cite{Ching:1994bd,Copi:2020qur,Copi:2022ire, Garoffolo:2022usx,Garoffolo:2023dsu,Kessler:2025idf}.  

In general, GWs appear to be a more promising probe of the tail effect cosmologically than do EMWs despite the fact that EMWs are more ubiquitously produced with high luminosity and are more readily detected over most wavebands.
One challenge is that EMWs of wavelength greater than about $100\;\mathrm{m}$ are excluded from the inner solar system by the plasma mass of the solar wind. 
Meanwhile, the tail effect arising from propagation through a perturbed nearly flat LFRW background  seems generally to be largest at wavelengths comparable to the geometric-curvature scales in the ``environment'' and scales $\lesssim 100\;\mathrm{m}$ are relatively rare, especially in electromagnetically transparent regions.
GWs sidestep all of these difficulties.

In \cite{Copi:2022ire}, some of us showed that 
relatively bright tail radiation should sometimes be induced by stars and stellar remnants in the black-hole-binary merger events visible to LVK.
This contrasted with earlier conclusions that the tail signal was too small to be detectable in the foreseeable future.  
Those investigations however had considered only two of the three epochs of the radiation tail.
These epochs can best be understood in the one-scattering approximation---where the tail is interpreted as the coherent scattering of waves, traveling at $c$, off the perturbations in the  geometry.
During the  ``early-time tail," there has not yet been enough time for a signal moving at $c$ to travel from the source to the surface of the physical perturber and thence to the observer---i.e., the perturber lies outside an ellipsoid of constant travel time (at $c$) with the source and the observer at its foci. 
For a Schwarzschild perturber, there is no detectable early time tail.
During the final ``late-time tail," when the perturber is inside that ellipsoid, there is  GW radiation, but it is very weak, and falls off quickly with time.

The new result of \cite{Copi:2022ire}, building on similar results for scalar and vector radiation in \cite{Copi:2020qur}, relates to a brief middle-time, when a wave traveling at $c$ from the wave source to the observer could have scattered off the geometry (not the matter!) from a point inside the physical perturber. 
During this short period, the amplitude of the radiation tail can be sizable. 
For a  weak-field spherically symmetric perturber on the midplane between the source and the observer and located outside its own Einstein ring, they were able to show that, to leading order in small quantities, the ratio of the second time derivative of the (transverse traceless) GW strain of the middle-time tail to that of the null-cone should be 
\begin{equation}
    \label{eqn:StoN}
    \varepsilon \equiv
    \frac{\vert\ddot{h}^{\mathrm{TT;middle}}_{ij}\vert}{\vert\ddot{h}^{\mathrm{TT;null}}_{ij}\vert} =  \frac{2G_N M_{P} f_P \ell  }{b^2 c^2} =\frac{f_P}{n_E^2}.
\end{equation}
Here $M_P$ is the mass of the perturber, $\ell$ is the perpendicular distance from the observer to the perturber plane, and $b$ is the distance of the perturber from the line-of-sight, so that $n_E>1$ is $b$ in units of the Einstein radius of the source-perturber-observer system. 
$f_P$ is a numerical factor that depends on the radial density profile of the perturber.

This middle-time tail is, again to leading order in various small quantities, a faithful (i.e. undistorted) echo of the null-cone signal, which we refer to as a ``glint."
The time delay between the null-cone signal and the middle-time tail is (for $b\ll \ell$)
\begin{widetext}
\begin{equation}
    \label{eqn:deltat}
    \Delta t \simeq \frac{b^2}{2 d c} \frac{d^2}{\ell(d-\ell)} \simeq n_E^2 \frac{M_P G_N}{c^3} \frac{d^2}{\ell(d-\ell)} \simeq \frac{f_P}{\varepsilon } \frac{M_P}{M_{\odot}} \frac{d^2}{\ell(d-\ell)} \times 10^{-5}\mathrm{s}\,,
\end{equation}
\end{widetext}
where $d$ is the line-of-sight distance from the source to the observer and $0 < \ell < d$, i.e. $\ell=d/2$ is the midplane for which \eqref{eqn:StoN} has been obtained.

If such perturbers are uniformly distributed in space then one can calculate the number of perturbers within $n_E$ Einstein radii of the line-of-sight between an observer and a GW source:
\begin{equation}
    \label{eqn:NofnE}
    N(n_E) = \frac{3}{2}  \Omega_P  (H_0 \ell  )^2 \frac{\Delta\ell}{\ell  } (n_E^2-1) .
\end{equation}
Here $\Omega_P$ is the fraction of the cosmological energy density in such perturbers, and $\Delta\ell/\ell $ is the fraction of the line of sight over which  \eqref{eqn:StoN} holds. 
Here we have assumed the pertuber lies along the mid-plane between the observer and the source.
Preliminary estimates are  \cite{Copi:2022ire} that $\Delta\ell/\ell \simeq 1$, i.e., that \eqref{eqn:StoN} holds over nearly the entire line of sight.

For perturbers that are typically about a solar mass, and GWs that are in the LIGO band, around a few 100~Hz, there will necessarily be a tradeoff between the relative amplitude $\varepsilon$ of the glint and the time delay.  Alternately, one can take $\ell(d-\ell)\ll d^2$ to boost the time delay, but at the price of a reduction in the event rate.  (Moreover, for $\ell\ll d$ and $d-\ell \ll d$, the results of \cite{Copi:2022ire} are not necessarily reliable.) 

In summary, given that for known perturbers with $M\simeq M_{\odot}$, $\Omega_P\ll1$ we do not expect the sample size of large SNR LVK events found in \cite{GWTC-3} to be sufficient to contain a detectable middle-time glint. 
Nevertheless, their existence motivates us to characterize the detectability of such glints in LVK data, with an eye toward much larger rates of detection of binary black hole merger events.






\section{Parameter Estimation for Gravitational-Wave Glints}
\label{sec:pe}

\subsection{Methods\label{sec:pe-methods}}
GW glints result in a relatively simple alteration to existing models for GWs from mergers of compact binaries.
For a given GW with associated null cone signal strain $h_{0 }(t)$ and Fourier transform $\mathcal{F}\{h_{0 }(t)\} \equiv \tilde{h}_{0}(f)$, the primary signal will be modified simply by the addition of an attenuated, spectrally equivalent copy, arriving at some time $\Delta t$ after the primary signal
\begin{align}
    \label{eq:glint-t}
    h(t) &= h_{0}(t) + \varepsilon h_{0}(t-\Delta t),
\end{align}
where $\varepsilon$ is the fractional amplitude of the primary signal, and $\Delta t$ is the arrival time difference between the primary signal and the tail.
In the frequency domain, its Fourier modes will be altered by
\begin{align}
    \label{eq:glint-f}
    \tilde{h}(f) &= \tilde{h}_{0}(f)[1 + \varepsilon \exp(2\pi i f \Delta t)].
\end{align}
Note that although the modification factor in Eq.~\ref{eq:glint-f} varies with frequency through its phase, the relative glint amplitude $\varepsilon$ is a frequency-independent constant, which means the glint is an undistorted copy of the primary signal, so the modification applies identically to any waveform model chosen for the primary signal.
While this parameterization is phenomenological, it can be mapped back to the geometric and physical parameters of the system.

\begin{figure*}
\centering
\begin{subfigure}{.5\textwidth}
  \centering
  \includegraphics[width=0.95\linewidth]{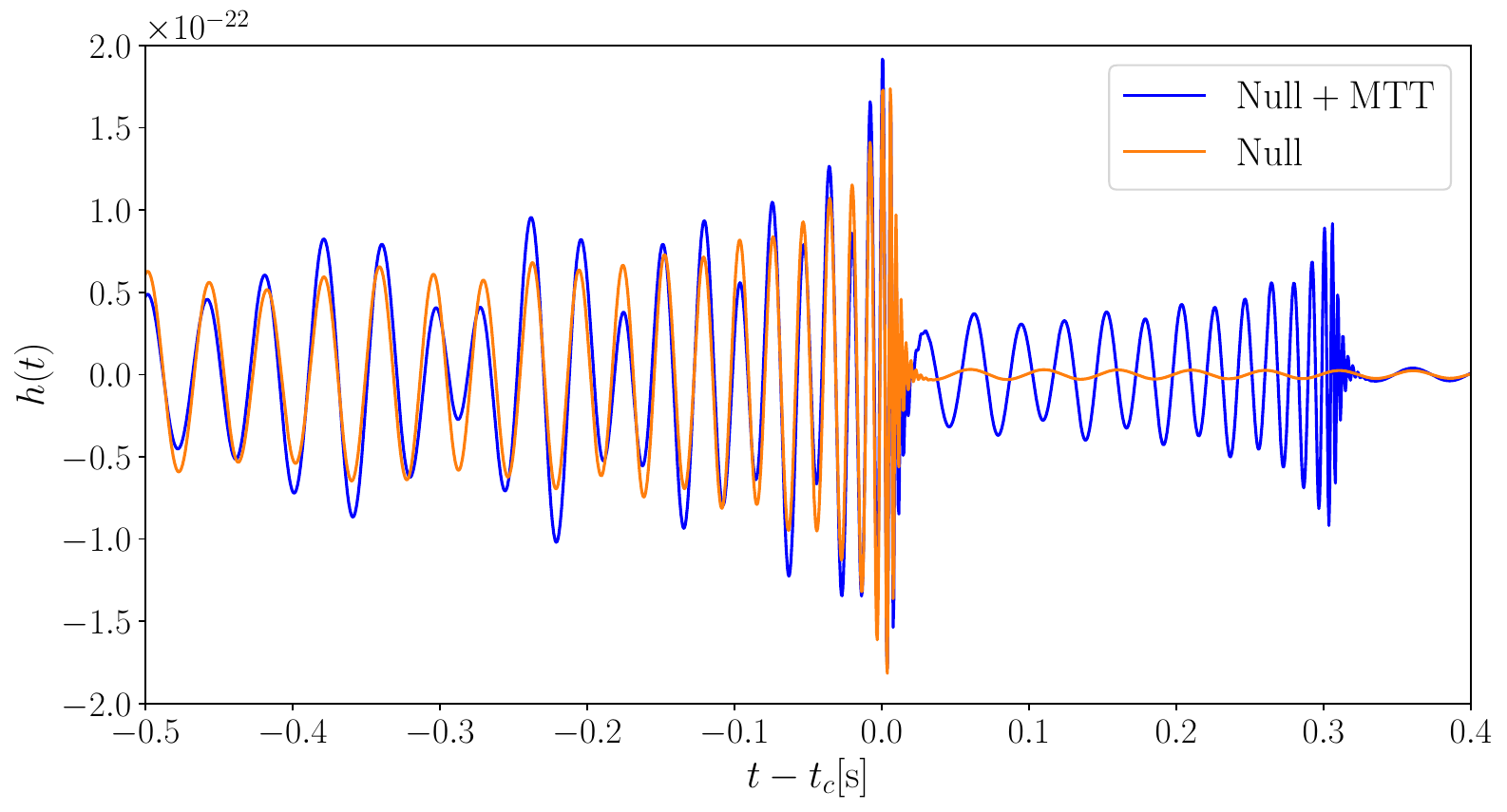}
  \caption{Time domain}
  \label{fig:TDecho}
\end{subfigure}%
\begin{subfigure}{.5\textwidth}
  \centering
  \includegraphics[width=0.95\linewidth]{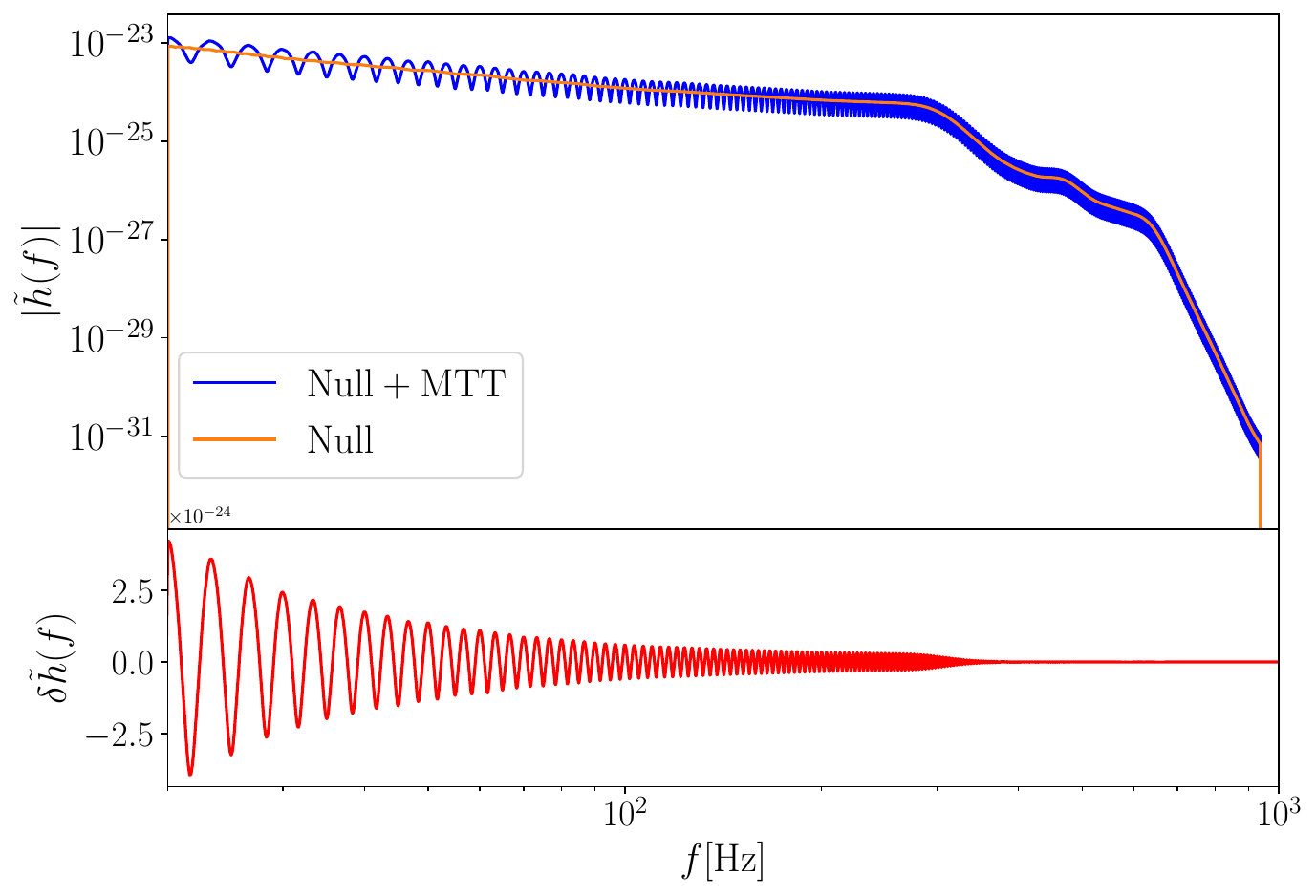}
  \caption{Frequency domain}
  \label{fig:FDecho}
\end{subfigure}
\caption{Detector response to a GW from a BBH merger accompanied by a middle-time tail (MTT) with relative amplitude $\varepsilon_{\mathrm{glint}} = 0.5$, arriving $\Delta t_{\mathrm{glint}}=0.3$ s after the primary signal.}
\label{fig:test}
\end{figure*}

Probabilistic information about the additional parameters can be extracted by employing Bayesian inference techniques.
For a model $m$ attempting to describe the signal in some data $d$, the probability density function on the model parameters $\boldsymbol{\theta}$ is given by the Bayesian posterior  
\begin{align}
    p(\boldsymbol{\theta}|d, m) &= \frac{p(\boldsymbol{\theta}|m)p(d|\boldsymbol{\theta}, m)}{p(d|m)},
\end{align}
which is calculated from the model prior $p(\boldsymbol{\theta}|m)$, the likelihood $p(d|\boldsymbol{\theta}, m)$, and the evidence $p(d|m)$.
The evidence $\mathcal{Z}\equiv p(d|m)$ normalizes the posterior by integration over the entire model parameter space:
\begin{align}
    \label{eq:evidence}
    \mathcal{Z} =  \int p(\boldsymbol{\theta}|m)p(d|\boldsymbol{\theta}, m) d\boldsymbol{\theta}. 
\end{align}

Comparison of the Bayesian evidence between two models provides a natural way to determine which model is preferred by the data.
The Bayesian posterior odds $\mathcal{O}_{12}$ that model $m_1$ is preferred over model $m_2$ can be related to the Bayes factor $\mathcal{B}_{12}$ in the following way:
\begin{align}
    \label{eq:odds}
    \mathcal{O}_{12} &= \frac{p(m_1|d)}{p(m_2|d)}
    =\frac{p(m_1)p(d|m_1)}{p(m_2)p(d|m_2)} 
    = \frac{p(m_1)}{p(m_2)}\mathcal{B}_{12}.
\end{align}
If both models are equally probable \textit{a priori}, then $p(m_1)=p(m_2)$ in Eq.~\ref{eq:odds}, and the odds reduces to the Bayes factor.
Since the Bayes factor is simply a ratio of the evidence for each model, determining which model is preferred by the data amounts to calculating the evidence for each model and finding their ratio.

In this work, we seek to determine whether any of the GW signals detected by the LIGO-Virgo network and cataloged in the cumulative GWTC-3~\cite{GWTC-1,GWTC-2,GWTC-3} contain GW glints.
Strong evidence for the existence of a glint will be finding a Bayes factor
\begin{equation}
    \label{eq:det-crit}
    \ln\mathcal{B}_{g0} \equiv \ln\left(\frac{\mathcal{Z}_{g}}{\mathcal{Z}_{0}}\right)>3,
\end{equation}
where $\mathcal{Z}_{g}=p(d|m_{g})$ is the evidence calculated using a GW model that includes a glint parameterized by $(\varepsilon,\Delta t)$ according to Eqs.~\ref{eq:glint-t} and \ref{eq:glint-f}, and $\mathcal{Z}_{0}=p(d|m_{0})$ is the evidence calculated using a standard GW model that does not include glint parameters.
The threshold of 3 in Eq.~\ref{eq:det-crit} is the criterion from Ref.~\cite{bayes-crit} for a {\it strong} preference for one model over another, in this case for a GW model with a glint over one without.
We used the Bilby inference software package \cite{bilby1,bilby2} with the Dynesty nested sampler \cite{nested-sampling,dynesty} to calculate evidences in all of our parameter estimation analyses.


As a proof of concept for the viability of a Bayes factor detection criterion for glints, we simulated 36 binary black hole (BBH) signals with the same null cone parameters but various values for $\varepsilon$ and $\Delta t$ and injected them into three simulated GW observatories representing the LIGO Hanford (H1), LIGO Livingston (L1) and Virgo (V1) observatories.
The simulated signals each had a total mass of $65~{\rm M}_\odot$ and aligned spin.
The glint parameters that we explored were:
\begin{eqnarray*}
\varepsilon&=&(0.01,0.2,0.4,0.6,0.8,1)\\
\Delta t&=&\left(\frac{1}{4f_c},\frac{1}{2.9f_c},\frac{1}{2f_c},\frac{1}{f_c},0.5~{\rm s},1~{\rm s}\right)
\end{eqnarray*}
where $f_c=66.67~{\rm Hz}$ is the approximate coalescence, or merger, frequency of our system.
The network SNR for each simulation was about 20. 
We performed Bayesian inference on each signal twice: once under the glint waveform model $m_{g}$ and once with a standard BBH waveform model $m_{0}$. 
To improve computational efficiency, we reduced the sampled parameter space by pinning the external system parameters to their injected values.
From the resulting Bayesian evidence from each analysis, we computed the Bayes factor between the two models.

We noticed several important features of glint analyses from this initial study.
First, we were able to observe a transition from negative log-Bayes factors ($\ln\mathcal{B}_{g0}<0$), where $m_0$ is preferred, to positive log-Bayes factors ($\ln\mathcal{B}_{g0}>0$), where $m_g$ is preferred, to log-Bayes factors much greater than 3, where $m_g$ is not merely preferred but strongly favored.
Second, when injections absent of a glint were analyzed with glint parameters included, we found no positive log-Bayes factors.
Third, we saw that the relative glint amplitude had a much stronger effect on the evidence than the time delay of the glint.
The time delay of the glint did not noticeably affect the evidence until the log-Bayes factor $\ln\mathcal{B}_{g0}$ became positive, at which point (generally) larger values of $\Delta t$ produced larger evidences.
However, this was mainly true for very large $\Delta t$; for $\Delta t<1/f_c$, the evidence values were typically comparable, with lower values sometimes resulting in slightly larger evidences.
Fourth, for signals whose log-Bayes factors exceeded 5, the glint parameters---especially $\Delta t$---were measured very precisely.
For signals with an SNR$\sim$20, we found sub-percent level $\Delta t$ precision and roughly $\sim$10\%-level $\varepsilon$ measurements.
Fifth, of the signals whose log-Bayes factors exceeded 5, in certain cases a shift in the one-dimensional marginal mass posteriors was present when glint parameters were excluded in parameter estimation.
This might mean that an observed GW event with a glint present but analyzed with a waveform model that does not include glint parameters could have a biased mass estimate.

The proof-of-concept simulations discussed so far were performed in so-called ``zero noise''.
A zero-noise analysis, which just means that the simulated data $d(t)$ is taken to be equal to the injected signal $s(t)$ since the noise $n(t)=0$, allows one to understand characteristic measurement uncertainties without the need to inject a signal into many distinct noise realizations.
Ref.~\cite{zero_noise_2} shows that Bayesian inference performed on a signal injected into zero noise is equivalent to averaging the posterior probability from that signal injected into many distinct noise realizations.
The overall statistical effect of noise is still incorporated in a zero-noise analysis through simulated power spectral densities, which are used to calculate likelihoods and network SNRs.
Thus, we incorporated the overall effect of noise without dealing with the statistical fluctuations and the computational cost of many noise realizations.
We concluded this proof-of-concept by injecting into just a handful of simulated noise instances.
We found that glint parameters were still accurately recovered and that none of the null signals that we analyzed were mistaken for glint signals due to individual noise realizations.

\subsection{Glint Model Comparison on the Gravitational Wave Transient Catalog}
\label{sec:pe_results}

The LVK published on 90 gravitational-wave detections during their first three observing runs \cite{GWTC-1,GWTC-2,GWTC-3}.
Since calculating Bayes factors is computationally expensive, involving two full Bayesian parameter estimation simulations for each considered event, we focus our search for glints in LVK's open data \cite{gwosc_1-2,gwosc_3} to loud signals far from Earth, because these will be the most likely to contain an observable glint.
Therefore, we only considered BBH and IMBH sources with ${\rm SNR}\ge 12$.
Under these criteria, we analyzed 34 events in GWTC-3 \cite{GWTC-3, gwosc_3}.
For our analysis, we used the waveform approximant IMRPhenomXPHM \cite{IMRPhenomXPHM} and priors consistent with those released in GWTC-3 \cite{GWTC-3} as well as a uniform prior in $\varepsilon=[0,1]$ and $\Delta t=[0,1]$~seconds.

We did not find strong evidence for the presence of a glint signature in any of the 33 total BBH events analyzed.
The BBH signals GW191109\_010717 and GW170608 were found with the largest Bayes factor: $\ln{\mathcal{B}_{g0}} = -0.8$.
Though this Bayes factor does not favor the presence of a glint, we include marginal posteriors for the recovered glint parameters for each event in Appendix~\ref{app_post} nonetheless.

Additionally, for the one IMBH signal that we analyzed, GW190521\_030229, we found $\ln\mathcal{B}_{g0}=5.2$, which indicates that our analysis very strongly favors a glint model over a model without glint parameters.
However, the further analysis of this signal suggests that this is a result of transient noise in the L1 detector as opposed to the presence of an actual glinted signal.
We provide a more robust analysis of GW190521\_030229 in Appendix~\ref{app_imbh}.

\section{Gravitational-Wave Glints in Modeled Searches}
\label{sec:search}

In addition to exploring all published GW candidates for evidence of a GW glint, we investigated whether current modeled searches for compact binary mergers provide sufficient sensitivity to GWs containing a glinted signal.
The LVK has several flagship compact binary merger search algorithms \cite{Nitz_2017, PhysRevD.90.082004, Alléné_2025, PhysRevD.95.042001, 2015arXiv150404632C}.  
All of these algorithms incorporate both SNR and a $\chi^2$ signal-consistency test as components of the final ranking statistic used to classify GW candidates \cite{PhysRevD.85.122006, PhysRevD.71.062001}. 
A signal containing a GW glint could affect each of these statistics differently, depending on the specific combination of glint parameters ($\varepsilon$, $\Delta t$) that is realized.
We investigated how the presence of a glinted GW signal affects the sensitivity of the flagship gstlal-inspiral compact binary search pipeline.
The results informed whether a new search through archival data using signal models that incorporate GW glints was warranted. We concluded that such a reanalysis was not necessary. However, we identified specific combinations of glint parameters that produce noticeable changes in search sensitivity, either enhancing or degrading the sensitivity depending on the specific glint parameters.
  
For this study, we used a template bank that covered the BBH and IMBH parameter spaces.  The BBH template bank consisted of 534,914 templates with total mass ranging from $6 \ M_\odot$ to $100 \ M_\odot$, mass ratio $q=m_1/m_2$ ranging from 1 to 10, and dimensionless spin parameter ranging from -1 to 1.  The IMBH template bank consisted of 85,220 templates with total mass ranging from $100 \ M_\odot$ to $758 \ M_\odot$, $q$ ranging from 1 to 10, and dimensionless spin parameter ranging from -1 to 1.  An additional set of 11,349 high-$q$ BBH templates were used with total mass ranging from $33 \ M_\odot$ to $400 \ M_\odot$, $q$ ranging from 10 to 50, and dimensionless spin parameter ranging from -1 to 1.


We generated a population of simulated GW signals from BBH and IMBH mergers.  The BBH set consisted of 26,665 signals generated with the IMRPhenomDpseudoFourPN waveform model, spanning component masses in the range $[5, 50] \ M_\odot$. The IMBH set included 26,664 signals generated with the SEOBNRv4pseudoFourPN waveform model, spanning component masses $[5, 300] \ M_\odot$.

We first performed a baseline search using the gstlal-inspiral pipeline with signals that did not include GW glints. This search was conducted over a 10-day stretch of data from the LVK third observing run (O3), from April 8, 2019 19:45:36 UTC (GPS 1238787954) to April 18, 2019 16:46:41 UTC (GPS 1239641219) \cite{KAGRA:2023pio}.
We then introduced GW glints into each simulated signal while keeping the template bank unchanged to assess the impact of glints on search sensitivity when using only non-glinted templates.
For each simulated signal, the relative amplitude $\varepsilon$ was drawn randomly from a uniform distribution between 0 and 1. The relative time delay $\Delta t$ was also drawn randomly between 0 and 1 seconds. In one simulation, $\Delta t$ was sampled from a uniform (linear) distribution, while in a second simulation it was sampled logarithmically, with a minimum value of $10^{-5}$ seconds.

\begin{figure*}[t]
    \centering
    \begin{subfigure}{0.48\linewidth}
        \centering
        \includegraphics[width=\linewidth]{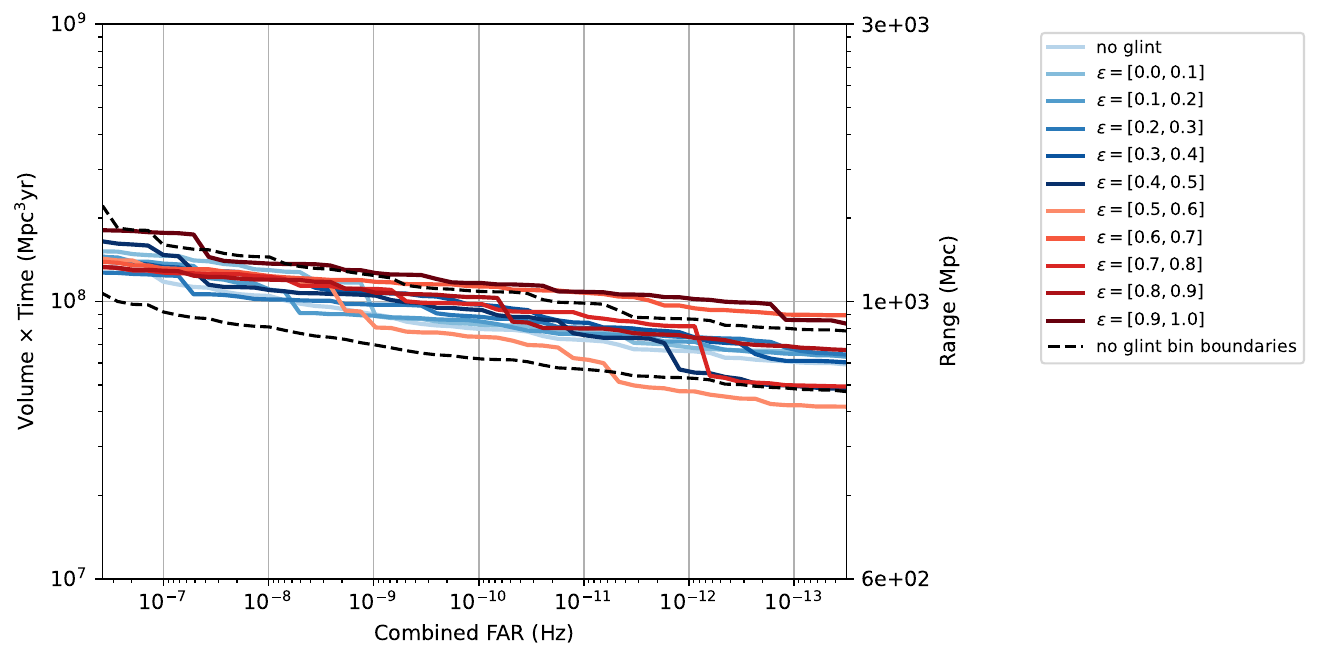}
    \end{subfigure}
    \hfill
    \begin{subfigure}{0.48\linewidth}
        \centering
        \includegraphics[width=\linewidth]{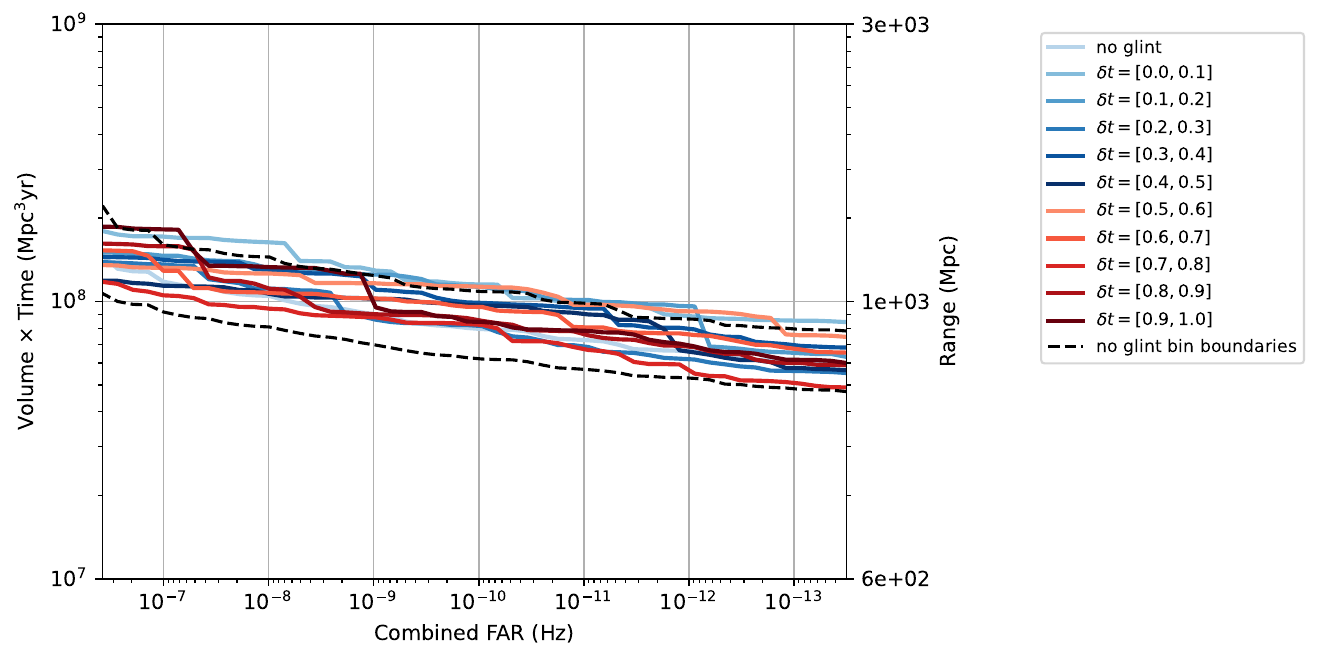}
    \end{subfigure}

    \medskip

    \begin{subfigure}{0.48\linewidth}
        \centering
        \includegraphics[width=\linewidth]{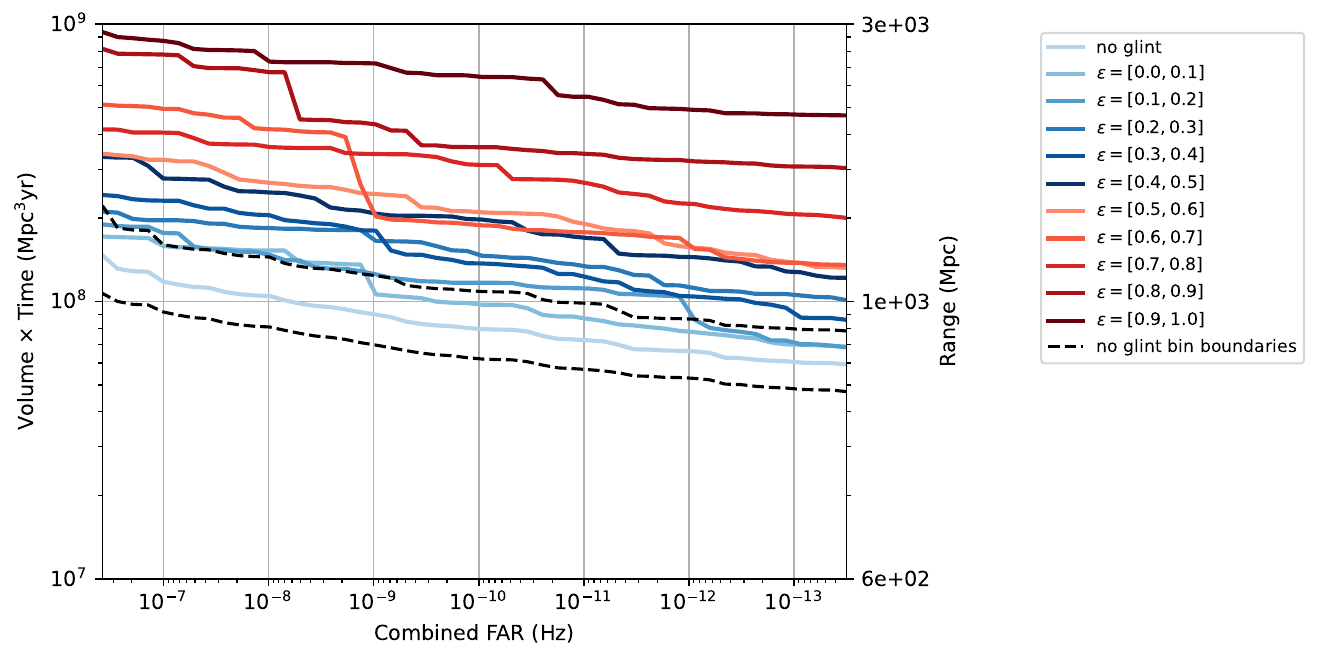}
    \end{subfigure}
    \hfill
    \begin{subfigure}{0.48\linewidth}
        \centering
        \includegraphics[width=\linewidth]{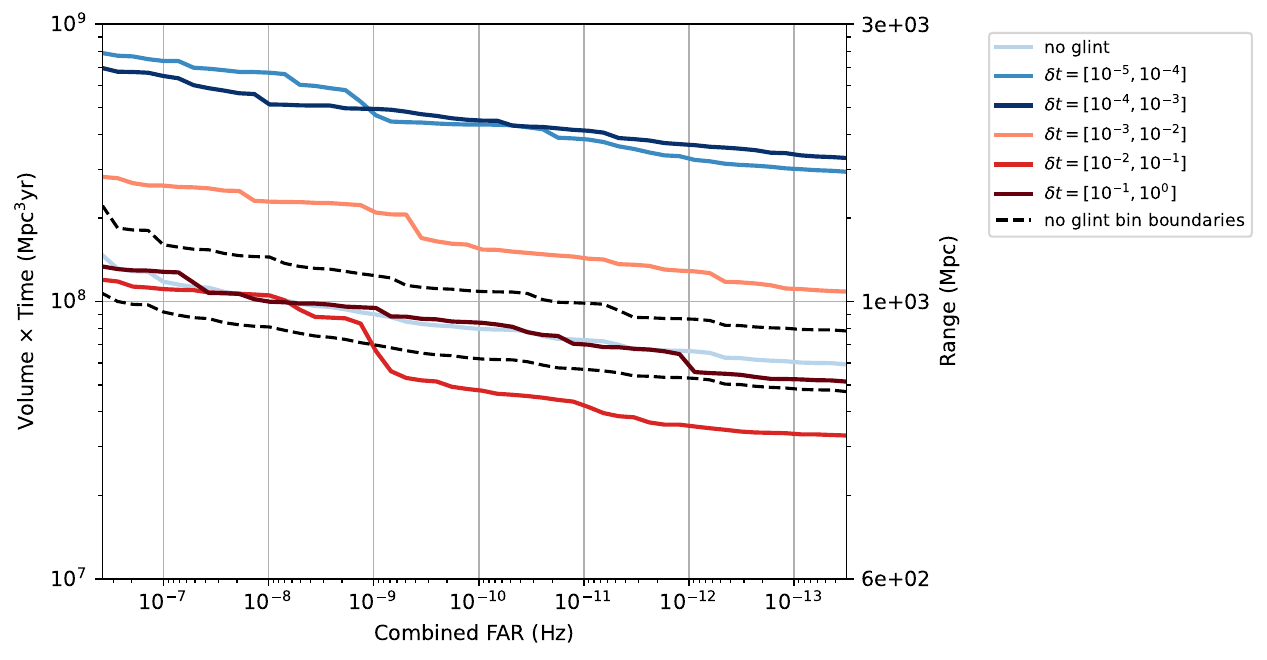}
    \end{subfigure}
    \caption{VT search sensitivity (left axis) or binary inspiral range (right axis) as a function of combined false alarm rate (FAR) for events with chirp masses between [4.5, 45] $M_\odot$ found in all detectors across the performed search.  The top row shows results for the simulated signals generated with uniform $\Delta t$ sampling and the bottom row shows results for the simulated signal set generated with logarithmic $\Delta t$ sampling.  The left column  shows sensitivity curves for individual bins in $\varepsilon$, and the right column shows sensitivity curves for individual bins in $\Delta t$.  Each plot contains one line for the search sensitivity resulting from the baseline simulated signal set that contained no glinted signals.  The dotted black lines on each plot show the bin boundaries for the non-glint baseline sensitivity.  When sampling logarithmically in $\Delta t$, which overemphasizes smaller $\Delta t$ values, there was a noticeable trend where search sensitivity was impacted, either enhanced or degraded, in a glint-parameter dependent way.}
    \label{fig:search_sensitivity}
\end{figure*}

A standard way to quantify the sensitivity of a compact binary coalescence search across parameter space is through quantification of the spacetime volume of the search, given by the three-dimensional spatial distance to a signal times the observing time of the data, as a function of false alarm rate across all observing detectors.  
Fig.~\ref{fig:search_sensitivity} shows the search sensitivity with different choices of binning for the glint parameters.  
For the simulated signals where both $\varepsilon$ and $\Delta t$ were uniformly drawn there is no noticeable difference between the no glint results and the results with GW glints added to each injection.
However, the simulated signals where the $\Delta t$ parameter was drawn logarithmically, which emphasizes smaller values of $\Delta t$, revealed an interesting corner of the parameter space where search sensitivity was either enhanced or degraded, depending on the value of $\Delta t$.  This interesting corner of parameter space is discussed in more detail in Appendix~\ref{app_search}.

In all other areas of parameter space, outside of large $\varepsilon$ and small $\Delta t$ corner of the glint parameter space, there was no statistically significant change in the search sensitivity when using simulated GWs that include glints, as can be seen in Fig.~\ref{fig:search_sensitivity}.
Therefore, we concluded that conducting a modeled search using templates that do not contain GW glint models would be unlikely to miss GW detections if a glint were present in the data.
As a result, we did not conduct an independent search of the LVK O1-O3 data using a template bank of signals that incorporates GW glints.
Instead, we analyzed the already published GW candidates found in GWTC-3 \cite{GWTC-3}, which was the most up-to-date catalog at the time of writing.


\section{Upper Limits}
\label{sec:pe_ul}


Since we found no strong evidence of GW glints in GWTC-3, we place upper limits on the probability of source-agnostic glints of a given relative strain, and, for the specific case of GW glints predicted by GR, on the cosmological abundance of the objects that would cause such glints.

To determine the upper limit $\varepsilon_{\rm ul}$ of the relative glint amplitude for a given event, we proceed as follows.
We first perform parameter estimation without including glint parameters in the model.
From this analysis, we identify the maximum-likelihood sample and use its parameters as the baseline waveform parameters. We then augment this set with the glint parameters, choosing $\varepsilon \in (0,1)$ and $\Delta t \in (0.1,0.2) \ \mathrm{s}$.\footnote{We found that the specific choice of $\Delta t$ did not noticeably affect the evidence calculations.}
We use these parameters to simulate a gravitational waveform with the waveform approximant IMRPhenomXPHM~\cite{IMRPhenomXPHM}.
We inject this waveform into zero noise using the detector PSDs near the time of the actual event.
We then use Bilby to estimate the evidence without glint parameters, $\mathcal{Z}_0$, and with glint parameters, $\mathcal{Z}_g$.
If the log-Bayes factor $\ln\mathcal{B}_{g0}<3$, we increase $\varepsilon$ and recompute $\ln\mathcal{B}_{g0}$; if $\ln\mathcal{B}_{g0}>3$, we decrease $\varepsilon$ and repeat the calculation.
We continue this process until we determine $\varepsilon_{\rm ul}$ to within $\pm 0.025$.
We refer to this procedure for estimating upper limits as the \textit{simulated-Bayes-factor method}.

\begin{figure*}[t] 
  \centering
  \begin{subfigure}{0.48\textwidth}
    \includegraphics[width=\linewidth]{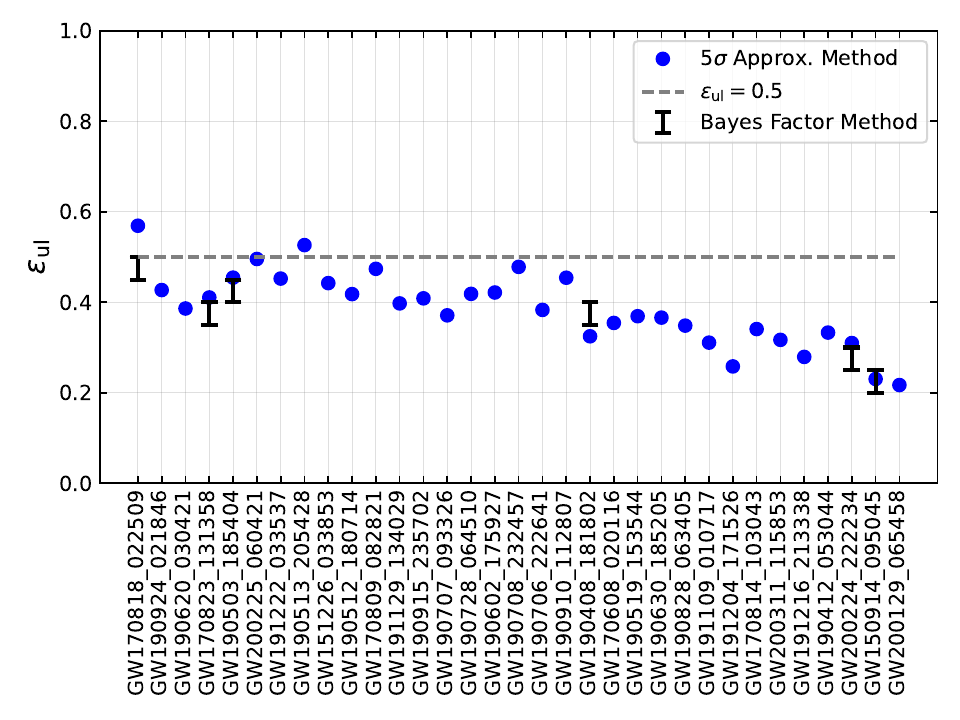}
  \end{subfigure}
  \hfill
  \begin{subfigure}{0.48\textwidth}
    \includegraphics[width=\linewidth]{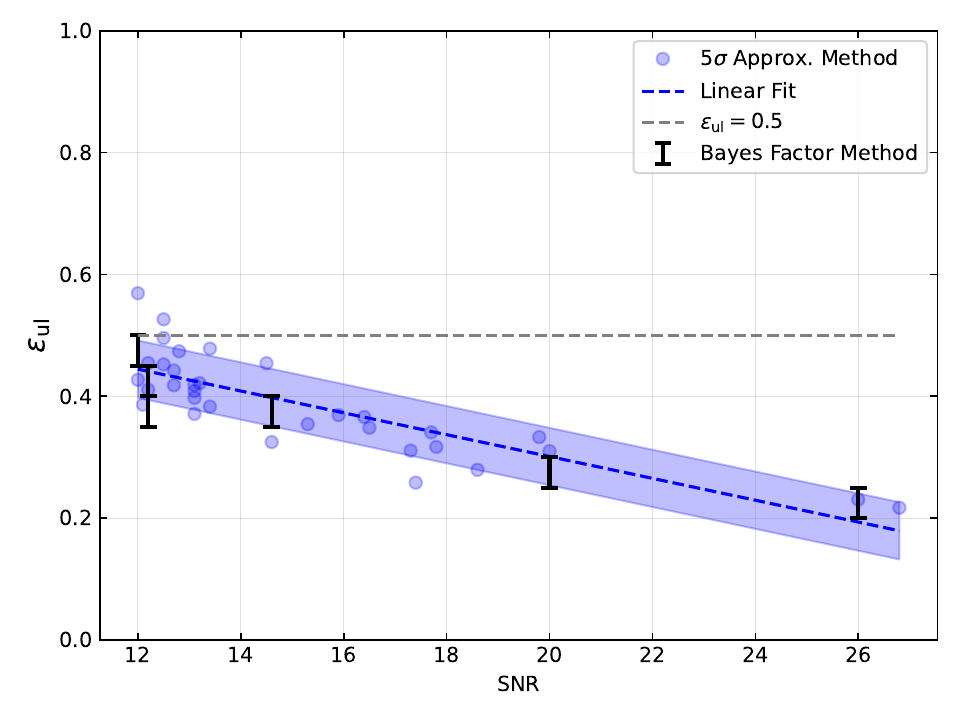}
  \end{subfigure}

  \caption{Plotted in the left panel are event-by-event upper limits on relative glint amplitude $\varepsilon_{\rm ul}$ for BBHs with ${\rm SNR}\ge12$ in GWTC-3. The blue dots were calculated using the $5\sigma$ approximation method, the black bars were calculated using the simulated-Bayes-factor method for a handful of events, and the dashed line is the conservative upper limit over all of these events, using the simulated-Bayes-factor method as the adopted upper limit for the largest $\varepsilon_{\rm ul}$.  The right panel presents the same plot but with SNR on the horizontal axis instead of event name. Additionally, we add a dashed linear fit to the $5\sigma$ approximation method estimates and shade the residual standard deviation of these estimates relative to the fit.} 
  \label{fig:UL}
\end{figure*}

Applying the simulated-Bayes-factor method to every analyzed event is computationally expensive. Instead, we estimate a conservative upper limit for all BBH events with ${\rm SNR} \ge 12$ by evaluating $\varepsilon_{\rm ul}$ for a small set of events with ${\rm SNR} \approx 12$ and adopting the largest resulting value as the overall upper limit.
This approach is justified because the upper limit on the relative glint amplitude, $\varepsilon_{\rm ul}$, depends on the significance of the event’s associated null signal. Since $\varepsilon$ is a relative parameter, the detectability of a glint scales with the event SNR. For example, if a glint with $\varepsilon = 0.2$ accompanies an event with ${\rm SNR} = 30$, the glint alone has an effective SNR of 6. In contrast, the same value of $\varepsilon$ for an event with ${\rm SNR} = 10$ corresponds to a glint SNR of only 2. The former would therefore be much more readily detectable than the latter.
As a result, the upper limits derived from the lowest-SNR events in our sample provide conservative bounds for all higher-SNR events. Across all BBH events with ${\rm SNR} \ge 12$, we find a conservative upper limit of $\varepsilon_{\rm ul} = 0.5$, shown in Fig.~\ref{fig:UL}.

\begin{figure}[ht]  
    \centering  
    \includegraphics[width=0.5\textwidth]{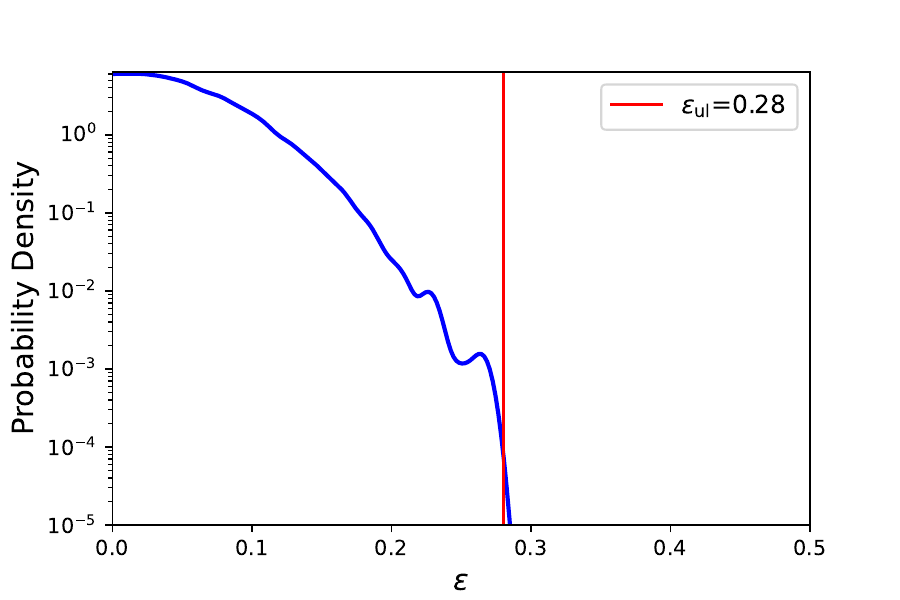}  
    \caption{Marginalized distribution for relative glint amplitude $\varepsilon$ for a simulated GW signal with the parameters of GW191216\_213338. In red is the upper limit of $\varepsilon$ using the $5\sigma$ approximation method described in Sec.~\ref{sec:pe_ul}.} 
    \label{fig:UL_example}  
\end{figure}

A less conservative but still computationally tractable way to place upper limits on $\varepsilon$ involves using an approximation method that requires just one parameter-estimation run per event.
By injecting a signal with no glint present in its waveform and running a parameter-estimation run where glint parameters are included in the waveform template, we recover a marginalized posterior density on $\varepsilon$ that peaks at $\varepsilon=0$ and eventually falls off for larger values of $\varepsilon$.
Fig.~\ref{fig:UL_example} is an example of such a distribution, smoothed with a Gaussian KDE, for a simulated GW signal with the parameters of GW191216\_213338.
Note that the probability density has negligible support for $\varepsilon\gtrsim0.3$, meaning that parameter estimation was not able to rule out the presence of a glint with $\varepsilon\lesssim0.3$.
Ref.~\cite{cornish_3sigma} used this insight to approximate upper limits on parameters by integrating their marginalized posterior distributions from zero to a probability of 99.7\% ($3\sigma$), and taking the parameter value at that probability bound to be the upper limit.

In this study, we originally calculated this $3\sigma$ approximation to $\varepsilon_{\rm ul}$.
However, we found that the $3\sigma$ threshold consistently underestimated $\varepsilon_{\rm ul}$ compared to the handful of events for which we used the simulated-Bayes-factor method to estimate $\varepsilon_{\rm ul}$.
We instead found that a $5\sigma$ threshold is more consistent with these estimates and can be used as a still somewhat conservative approximation for event-by-event upper limits on $\varepsilon$.
We additionally exclude samples for which $\Delta t$ is very close to zero, since in this limit $\varepsilon$ becomes degenerate with luminosity distance, allowing it to take on nearly any value.

We refer to this procedure for estimating upper limits as the $\mathit{5\sigma}$~{\it approximation method}.
These results are presented in Fig.~\ref{fig:UL}.
Because the $\varepsilon_{\rm ul}$ values in the $5\sigma$ approximation method are estimated from the tails of probability distributions, there is some uncertainty in each estimate.
Despite this, we observe an approximately linear trend in the upper limits as a function of SNR, and we show a corresponding linear fit in the right panel of Fig.~\ref{fig:UL}.
We characterize the uncertainty of the measurements made using the $5\sigma$ approximation method with a residual standard deviation of this fit plotted in the shaded region.
This shaded region appears consistent with the handful of measurements using the simulated-Bayes-factor method.

\begin{table}[ht]
\begin{tabular}{|c|c|c|c|c|}
\hline
 \hspace{9mm} Event \hspace{9mm} & \hspace{1mm} $\varepsilon_{\mathrm{ul}}$ \hspace{1mm} & $N(\leq \varepsilon_{\mathrm{ul}})$ & \hspace{0.1mm} SNR \hspace{0.1mm} & d (Gpc)\\
\hline
GW200129\_065458 & 0.22 &  1 & 26.8 & 0.89\\
GW150914\_095045 & 0.23 &  2 & 26.0 & 0.47\\
GW191204\_171526 & 0.26 &  3 & 17.4 & 0.64\\
GW191216\_213338 & 0.28 &  4 & 18.6 & 0.34\\
GW200224\_222234 & 0.31 &  5 & 20.0 & 1.71\\
GW191109\_010717 & 0.31 &  6 & 17.3 & 1.29\\
GW200311\_115853 & 0.32 &  7 & 17.8 & 1.17\\
GW190408\_181802 & 0.32 &  8 & 14.6 & 1.54\\
GW190412\_053044 & 0.33 &  9 & 19.8 & 0.72\\
GW170814\_103043 & 0.34 & 10 & 17.7 & 0.61\\
GW190828\_063405 & 0.35 & 11 & 16.5 & 2.07\\
GW170608\_020116 & 0.35 & 12 & 15.3 & 0.34\\
GW190630\_185205 & 0.37 & 13 & 15.6 & 0.87\\
GW190519\_153544 & 0.37 & 14 & 15.9 & 2.60\\
GW190707\_093326 & 0.37 & 15 & 13.1 & 0.85\\
GW190706\_222641 & 0.38 & 16 & 13.4 & 3.63\\
GW190620\_030421 & 0.39 & 17 & 12.1 & 2.91\\
GW191129\_134029 & 0.40 & 18 & 13.1 & 0.79\\
GW190915\_235702 & 0.40 & 19 & 13.1 & 1.75\\
GW170823\_131358 & 0.41 & 20 & 12.2 & 1.97\\
GW190512\_180714 & 0.42 & 21 & 12.7 & 1.46\\
GW190728\_064510 & 0.42 & 22 & 13.1 & 0.88\\
GW190602\_175927 & 0.42 & 23 & 13.2 & 2.84\\
GW190924\_021846 & 0.43 & 24 & 12.0 & 0.55\\
GW151226\_033853 & 0.44 & 25 & 12.7 & 0.46\\
GW191222\_033537 & 0.45 & 26 & 12.5 & 3.00\\
GW190910\_112807 & 0.45 & 27 & 14.5 & 1.52\\
GW190503\_185404 & 0.45 & 28 & 12.2 & 1.52\\
GW170809\_082821 & 0.47 & 29 & 12.8 & 1.07\\
GW190708\_232457 & 0.48 & 30 & 13.4 & 0.93\\
GW200225\_060421 & 0.50 & 31 & 12.5 & 1.15\\
GW190513\_205428 & 0.53 & 32 & 12.5 & 2.21\\
GW170818\_022509 & 0.57 & 33 & 12.0 & 1.08\\
\hline
\end{tabular}
\caption{Binary black hole merger events with SNR~$\geq12$ in increasing order of the upper-limit on $\varepsilon$ using the $5\sigma$ approximation method. SNR and distances are from the third GW transient catalog GWTC-3~\cite{Nitz:2021zwj, GWTC-3}.}
\label{tab:events}
\end{table}

\begin{figure}
    \centering
    \includegraphics[width=.9\columnwidth]{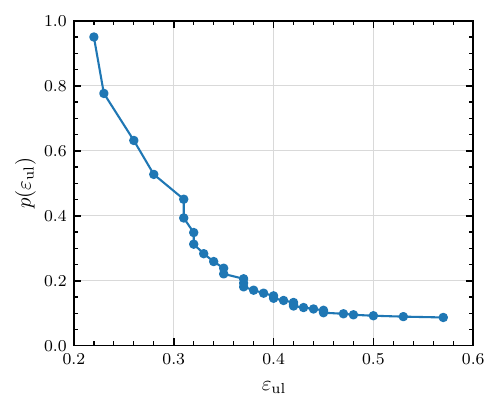}
    \caption{95\% C.L. upper limit on the probability of a glint as a function of $\varepsilon$ assuming equal probability for each binary black-hole merger event.}
    \label{fig:permergerlimit}
\end{figure}

\begin{figure}
    \centering
    \includegraphics[width=.9\columnwidth]{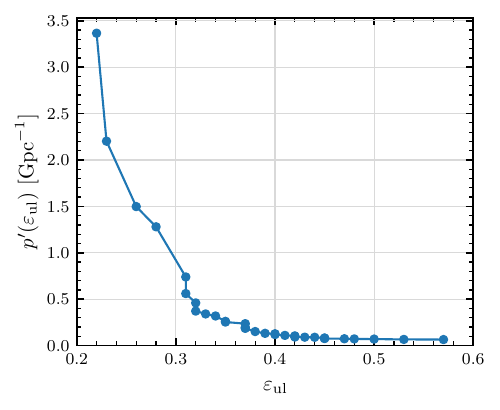}
    \caption{95\% C.L. upper limit on the expected number of glints per Gpc along the line of sight between the observer and the merger as a function of $\varepsilon$ assuming that rate is constant. }
    \label{fig:perGpclimit}
\end{figure}

\begin{figure}
    \centering
    \includegraphics[width=.9\columnwidth]{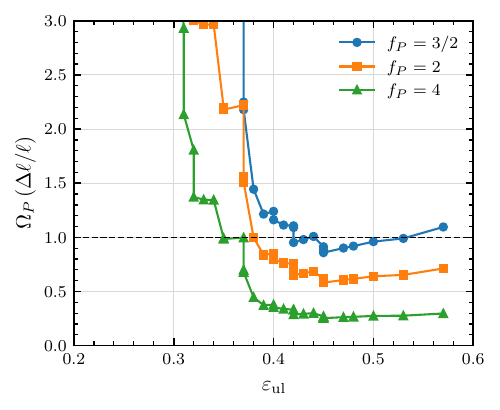}
    \caption{95\% C.L. upper limit on $\Omega_p (\Delta\ell/\ell)$ as a function of $\varepsilon$ for three representative values of $f_P$.  
    Populations that lie above the curves are ruled out. 
    The maximum possible value of $(\Delta\ell/\ell)$ is 1. 
    Known objects such as stars and compact stellar remnants are already known to have much smaller values of $\Omega_p$. 
    \label{fig:perOmegaplimit}
    }
\end{figure}

From the event-by-event upper limits $\varepsilon_{\rm ul}$ presented in Fig.~\ref{fig:UL} and tabulated in Tab.~\ref{tab:events}, we can estimate an upper limit on the rate at which detectable glints occur in the LVK detector network.
We adopt three approaches to obtaining an upper limit on glints.
In the first approach, we assume that the probability of a merger event yielding a glint of relative amplitude $\varepsilon$ above some threshold value and with $\Delta t$ in the search range ($0\leq \Delta t \leq 1 $)  is independent of other observable properties of the merger event. 
If the probability of such a glint is $p$, and 
if there are $N(\leq\varepsilon_{\mathrm{ul}})$ observed merger events for which a glint with $\varepsilon>\varepsilon_{\mathrm{ul}}$ would have been detected, then the probability of no glints being observed is 
\begin{equation}
    \bar{\mathcal{P}} =  (1-p)^{N(\varepsilon_{\mathrm{ul}})}.
\end{equation}
Requiring that $\bar{\mathcal{P}}\geq 0.05$, we find an upper limit on $p$ as a function of $\varepsilon$
\begin{equation}
    \label{eqn:plimit_perBBHevent}
    p(\varepsilon_{\mathrm{ul}}) \leq 1-0.05^{1/N(\leq\varepsilon_{\mathrm{ul}})}.
\end{equation}

In Fig.~\ref{fig:permergerlimit}, we plot the upper limit from Eq.~\eqref{eqn:plimit_perBBHevent}
on the per-merger probability $p$ of a glint as a function of the threshold $\varepsilon_{\mathrm{ul}}$ assuming constant value of $p$ for all $\varepsilon\geq\varepsilon_{\mathrm{ul}}$.
We see that for strong glints the limit approaches $p(\varepsilon)\lesssim0.1$.
In other words, not more than 10\% of BBH events can be expected to have a glint with $\varepsilon\gtrsim0.5$ for $0\le\Delta t \le 1$ if the probability per event is a constant for events with $\varepsilon\gtrsim0.25$.

In the second approach we assume that the expected number of glints (above a threshold $\varepsilon$) is proportional to the distance along the line of sight between the BBH event and the observer (and independent of all other properties of the event).
In Tab.~\ref{tab:events}, we list the distance $d_i$ (in Gpc) inferred for each of the 33 BBH merger events with SNR~$\geq12$ that we considered.

Denoting the probability per unit length of glint production $p'$,
the probability of no glint being observed with $\varepsilon\geq\varepsilon_{\mathrm{ul}}$ is
\begin{equation}
    \bar{\mathcal{P}} = e^{- p' \sum_{i=1}^{N(\leq\varepsilon)}d_i}\,.
\end{equation}
Again requiring that $\bar{\mathcal{P}}\geq 0.05$ we have
\begin{equation}
    p'(\varepsilon_{\mathrm{ul}}) \leq \frac{-\ln{0.05}}{\sum_{i=1}^{N(\leq\varepsilon)}d_i}
\end{equation}
In Fig.~\ref{fig:perGpclimit}, we plot the upper limit on the probability per unit distance $p'$ of a glint as a function of the threshold $\varepsilon$.

Our third approach is more specific to the GW glints predicted by GR described in Sec.~\ref{sec:tails}. 
Substituting Eq. \ref{eqn:StoN} for $\varepsilon$ as a function of $n_E^2$ into Eq.~\ref{eqn:NofnE},
the  expected number of glints with relative strain of at least $\varepsilon$ that will be observed in association with a particular BBH is 
\begin{equation}
    N(n_E) = \frac{3}{2}  \Omega_P  (H_0 \ell  )^2 \frac{\Delta\ell}{\ell  } \left(\frac{f_P}{\varepsilon}-1\right) .
\end{equation}
The number of expected glints for each BBH event with $\varepsilon$ above a threshold value is proportional to the square of the distance between us and the perturber.
Since  $\Delta\ell/\ell\leq1$,
the failure to observe any glints  can be used to establish a limit on $\Omega_P$, the fraction of the cosmological critical density in perturbers that would cause glints within the range of detectable $\Delta t$. 
This is shown in Fig.~\ref{fig:perOmegaplimit} for three representative values of $f_P$. 

Unfortunately, the upper limits on $\Omega_P$ are far larger than the expected value of $\Omega$ for, e.g., stars and compact stellar remnants.  
While these limits are of interest for non-black-hole macroscopic dark matter candidates \cite{Witten:1984rs,Zhitnitsky:2006x,Jacobs:2014yca,Bramante:2026wzh}, such candidates predict $\Delta t$ far below the range to which LVK is currently sensitive.

\section{Discussion}
\label{sec:discussion}

We find no evidence that a glinted GW model is favored over an unglinted GW model for any of the analyzed BBH or IMBH detections with $\mathrm{SNR} \ge 12$, beyond what can be attributed to noise. We therefore claim no detection of GW glints in LVK data from the first through the third observing runs.
We also show that this nondetection is not due to existing LVK searches being insensitive to glinted signals. If a glint with sufficient relative amplitude and an appropriate time delay had accompanied any of the detected events, a standard compact binary coalescence search would likely still have identified the signal, even when using templates that do not explicitly include glints.
Finally, we use our upper limits on the relative glint amplitude, $\varepsilon$, to place constraints on the properties of the perturber populations that could produce such signals.

It is not surprising that our analysis did not confidently identify any glint signals. The sensitivity of the LVK detectors during the first three observing runs requires relatively large values of $\varepsilon$ for a glinted signal to be detectable.
In Ref.~\cite{Copi:2022ire}, the authors predicted that a glinted signal with $\varepsilon = 0.3$ would be expected only once every 225 events. According to Fig.~\ref{fig:UL}, glints of this strength would be detectable only for signals with $\mathrm{SNR} \gtrsim 20$, of which there are very few in the first three observing runs. Our analysis was therefore unlikely to detect such a signal.
Furthermore, the expected time delays for glints produced by known astrophysical populations -- or even by potential macroscopic dark matter populations -- are much shorter than the delays to which the current search is sensitive.

However, the improved sensitivity of the LVK detectors during O4a alone has already provided a roughly equal number of BBH signals with $\rm{SNR}\ge 12$ as the first three observing runs.
At the conclusion of the fourth observing run, the LVK will have accumulated a significantly larger number of high-SNR signals than what was analyzed in this study.
Future glint searches will thus have a higher likelihood of detecting a gravitational glint or else result in tighter constraints on glint amplitudes and perturber properties.

\section{Acknowledgments}

This material is based upon work supported by NSF's LIGO Laboratory which is a major facility fully funded by the National Science Foundation (NSF).
LIGO was constructed by the California Institute of Technology and Massachusetts Institute of Technology with funding from the United States NSF, and operates under cooperative agreement PHY–2309200. Advanced LIGO was built under award PHY–0823459.
The authors gratefully acknowledge the support of the United States NSF for the construction and operation of the LIGO Laboratory and Advanced LIGO as well as the Science
and Technology Facilities Council (STFC) of the United Kingdom, the Max-Planck-Society
(MPS), and the State of Niedersachsen/Germany for support of the construction of Advanced LIGO and construction and operation of the GEO600 detector. 
Additional support for Advanced LIGO was provided by the Australian Research Council (ARC).
The authors are grateful for computational resources provided by the LIGO Laboratory and supported by NSF Grants PHY-0757058 and PHY-0823459.
KK, EB, MC, JB, MVK, LW, and MW are supported by NSF grant PHY-2308796.  
GDS was supported for work on this manuscript by DOE grant DESC0009946.





\clearpage
\appendix

\section{Impact of small $\Delta t$ and large $\varepsilon$ glints on search sensitive volume-time}
\label{app_search}
\begin{figure*}[htbp]
    \centering
    \begin{subfigure}{0.45\textwidth}
        \centering
        \includegraphics[width=\linewidth]{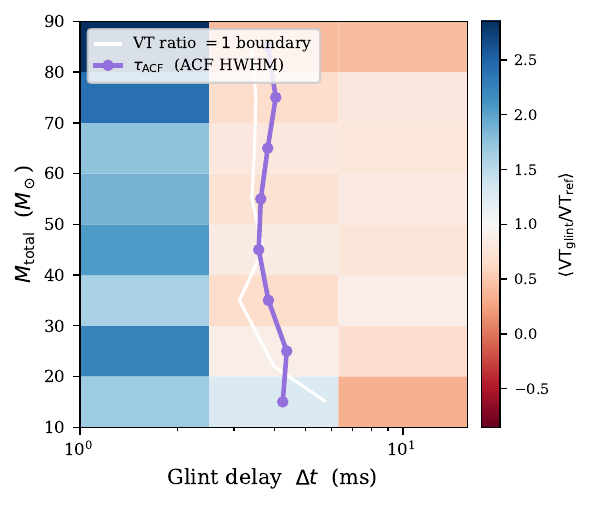}
        \caption{}
        \label{fig:fig1}
    \end{subfigure}
    \hfill
    \begin{subfigure}{0.45\textwidth}
        \centering
        \includegraphics[width=\linewidth]{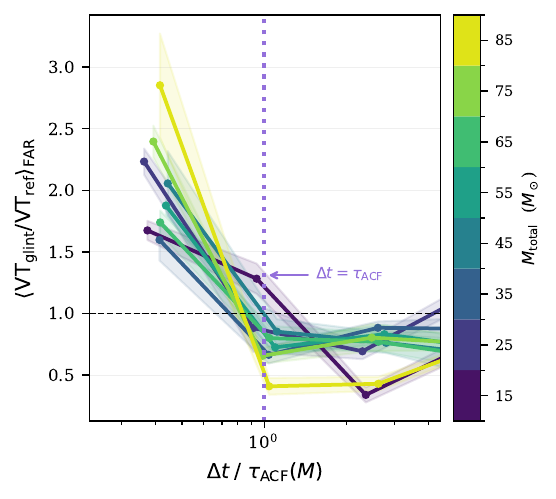}
        \caption{}
        \label{fig:fig2}
    \end{subfigure}
    
    \caption{Gain-to-loss transition in matched-filter sensitivity for gravitational-wave glints. Left:            
  FAR-averaged ratio of sensitive volume–time $\langle \mathrm{VT}_{\rm glint} / \mathrm{VT}_{\rm ref}    
  \rangle$ as a function of glint delay $\Delta t$ and binary total mass $M_{\rm total}$, for       
  large glint amplitudes ($\varepsilon \in [0.7, 1.0]$). Blue (red) indicates enhanced (degraded)
  sensitivity relative to the no-glint baseline. The white contour marks the gain-to-loss boundary      
  (ratio $= 1$). The purple curve shows the matched-filter autocorrelation half-width $\tau_{\rm ACF}$. Right: The same VT ratio curves for each mass bin (colored by $M_{\rm
  total}$) with the $\Delta t$ axis rescaled by $\tau_{\rm ACF}(M)$. The transition collapses to a
  common crossing near $\Delta t / \tau_{\rm ACF} = 1$ (vertical dotted line) across all mass bins. This
   demonstrates that $\tau_{\rm ACF}$ sets the characteristic timescale for the glint delay $\Delta t$ that leads to the search sensitivity cross-over from sensitivity gain to sensitivity loss, independent of total mass.}
    \label{fig:vt_transition}
\end{figure*}

When investigating the impact of GW glints on the sensitivity of modeled searches for compact binary mergers, we determined that GW glints with large relative amplitudes $\varepsilon$ and small time delays $\Delta t$ would have a noticeable impact on search volume-time (VT) sensitivity.
In order to better understand the interesting corner of glint parameter space, we narrowed in on simulated signals with large values of $\varepsilon=[0.7, 1.0]$ and values of $\Delta t$ that spanned the apparently most interesting region from $[10^{-2}, 10^{-1}]$~s.
Our results show that small values of $\Delta t$ lead to enhanced search sensitivity, and as $\Delta t$ increases, there is a transition point where the glinted GW signal degrades the overall search sensitivity.  These results can be seen in Fig.~\ref{fig:vt_transition}.

At small values of $\Delta t$, the glinted signal adds coherently to the SNR without significantly impacting the $\chi^2$ signal-consistency test. 
However, as $\Delta t$ grows, the glinted signal begins to add a second peak to the SNR time series, which in turn degrades the $\chi^2$ test.  
The glint time delays that span the transition from sensitivity gain to sensitivity loss due to the glinted signal was found to be consistent with the width (in time) of the autocorrelation function for a given template signal.
For each template, the SNR time series has an autocorrelation function (ACF) given by  
\begin{equation}
\text{ACF}(\tau) \propto \int_0^\infty \frac{|\tilde{h}(f)|^2}{S_n(f)} e^{2\pi i f \tau} df
\end{equation}
where $\tilde{h}(f)$ is the template waveform and $S_n(f)$ is the one-sided noise power spectral density. 
The ACF peak width (in time) $\tau_\mathrm{ACF}$ -- defined here as the half-width at half-maximum of $|\mathrm{ACF}(\tau)|$ -- characterizes the relevant time scale for glinted signals to begin to degrade search sensitivity: two signals separated by $\Delta t \ll \tau_\mathrm{ACF}$ produce nearly identical SNR time series, while a glint at $\Delta t \gtrsim \tau_\mathrm{ACF}$ falls outside the coherent peak and begins to register as a separate event. 
The values of $\tau_{\rm ACF}$ for each mass bin in Fig.~\ref{fig:vt_transition} were calculated as the median $\tau_{\rm ACF}$ for all templates in a given mass bin.
The plots in Fig.~\ref{fig:vt_transition} show that $\tau_{\rm ACF}$ provides roughly the correct $\Delta t$-scaling for the transition from enhanced search sensitivity due to the presence of a glinted signal to decreased search sensitivity due to the presence of a glinted signal.

\section{GW191109\_010717 and GW170608 glint posteriors}
\label{app_post}
\begin{figure}[h!]  
    \centering  
    \includegraphics[width=0.5\textwidth]{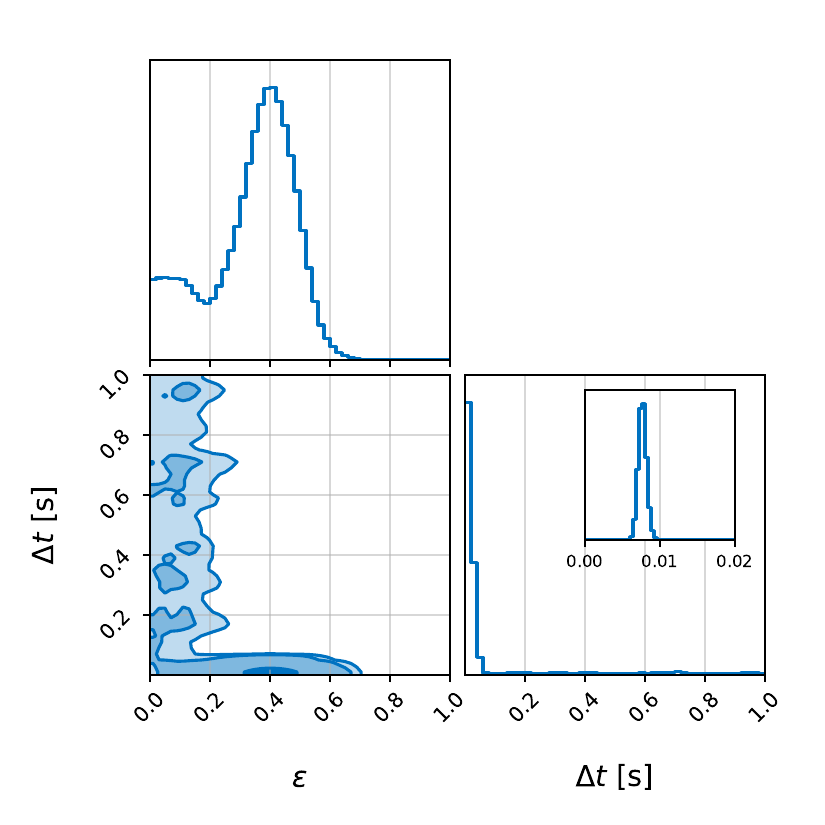}
    \caption{GW191109\_010717 corner plot of the marginalized posterior probabilities on the glint parameters $\varepsilon$ and $\Delta t$. The inset plot in $\Delta t$ zooms in to show the peak in $\Delta t$ more clearly.}
    \label{fig:Corner_Plot_191109}  
\end{figure}

\begin{figure}[h!]  
    \centering  
    \includegraphics[width=0.5\textwidth]{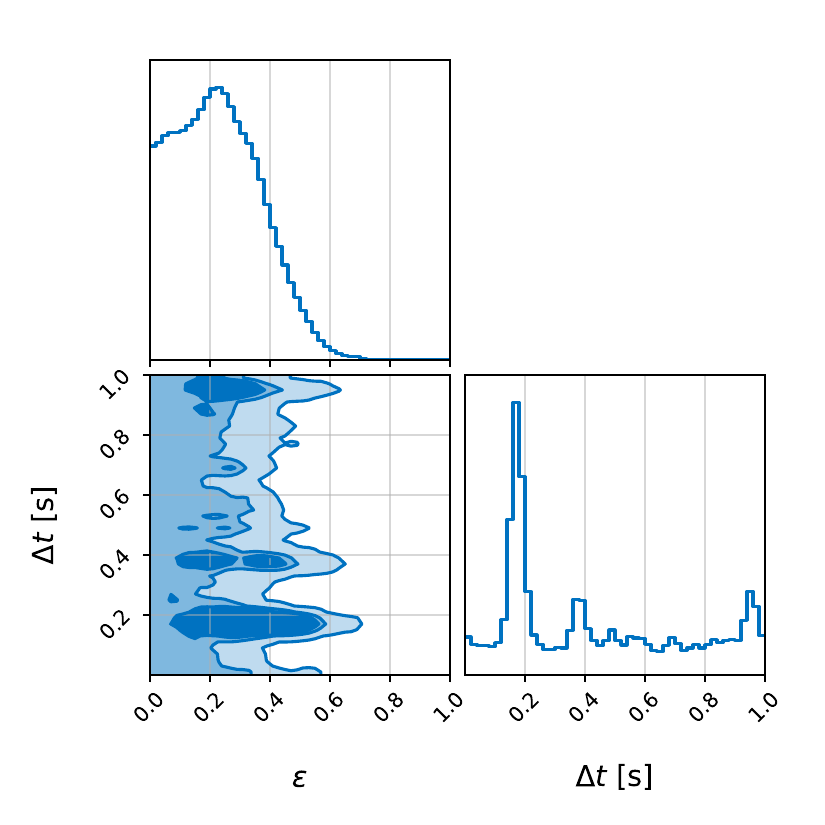}
    \caption{GW170618 corner plot of the marginalized posterior probabilities on the glint parameters $\varepsilon$ and $\Delta t$.}
    \label{fig:Corner_Plot_170608}  
\end{figure}

Figs.~\ref{fig:Corner_Plot_191109} and~\ref{fig:Corner_Plot_170608} present the marginal posterior distributions of the glint parameters for GW191109\_010717 and GW170618, which are the BBH signals with the largest $\ln\mathcal{B}_{g0}$, as presented in Sec.~\ref{sec:pe_results}.
The marginal posteriors notably peak at values greater than zero in both $\Delta t$ and $\varepsilon$.
This indicates that, although the Bayes factor does not meet our detection threshold, the maximum-posterior region of parameter space corresponds to a waveform with a nonzero glint component, with $\varepsilon\sim0.4$ for GW191109\_010717 and $\varepsilon\sim0.2$ for GW170608.
This behavior reflects the structure of Bayesian model comparison: models with additional parameters are penalized through the Occam factor, which disfavors unnecessarily complex models unless the improvement in likelihood is sufficiently strong.
In the case of GW191109\_010717 and GW170608, although a glinted waveform provides a slightly better fit to the data, this improvement is not sufficient to overcome the Occam penalty. As a result, the overall evidence remains lower than that of the simpler (unglinted) model, yielding $\ln\mathcal{B}_{g0} < 0$.

We also note that the evidence is impacted by the size of our prior choice.
While the $\varepsilon$ prior is physically restricted to $\varepsilon=[0,1]$, there is no known physical bound on $\Delta t$.
We chose to set our $\Delta t$ prior to $\Delta t=[0,1]$~seconds to be sufficiently large to catch glints from a wide range of theorized sources and sufficiently small to reach convergence in our sampling.
We found that a prior on $\Delta t$ much wider than 1~second required more live points than we were able to analyze practically given our computational resources.

\section{Further analysis of GW190521\_030229}
    \label{app_imbh}
\begin{figure*}[t] 
  \centering
  \begin{subfigure}{0.48\textwidth}
    \includegraphics[width=\linewidth]{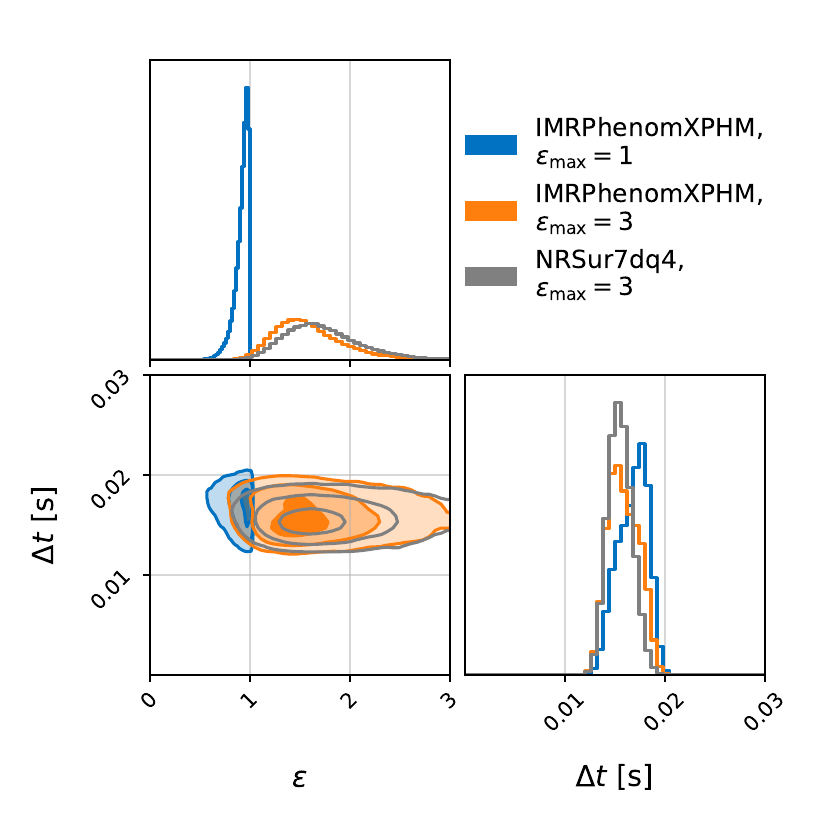}
  \end{subfigure}
  \hfill
  \begin{subfigure}{0.48\textwidth}
    \includegraphics[width=\linewidth]{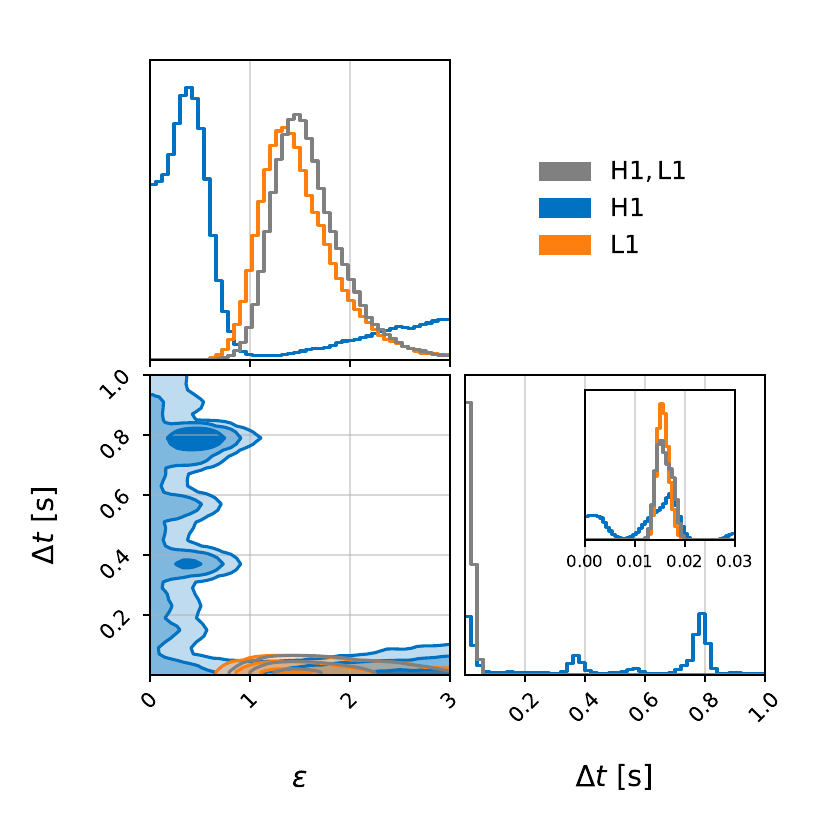}
  \end{subfigure}

  \caption{GW190521\_030229 corner plots of the marginalized posterior probabilities on the glint parameters $\varepsilon$ and $\Delta t$ using various priors, waveform approximants, and data streams. The left panel compares the marginalized posteriors using a uniform prior on relative glint amplitude $\varepsilon$ between $[0,\varepsilon_{\rm max}]$, where $\varepsilon_{\rm max}$ is noted in the legend, and the two waveform approximants also noted in the legend. The right panel compares the marginalized posteriors using data from just the Hanford detector (H1), just the Livingston detector (L1), and both the Hanford and Livingston detectors (H1, L1).}
  \label{fig:GW190521-followup}
\end{figure*}

As stated in Sec.~\ref{sec:pe_results}, when we analyzed GW190521\_030229, we recovered $\ln\mathcal{B}_{g0}=5.2$, indicating that the glint model is very strongly favored relative to the model without a glint.
However, upon further inspection, the marginalized posteriors rail against the $\varepsilon$ prior at $\varepsilon=1.0$ and peak at $\Delta t = 0.015$~seconds, as seen in the left panel of Fig.~\ref{fig:GW190521-followup} in blue.

We followed up by expanding the $\varepsilon$ prior to a maximum of $\varepsilon = 3.0$ to test whether this initial marginalized posterior was indicative of the $\varepsilon$ recovery peaking at $\varepsilon=1.0$ or at $\varepsilon>1.0$.
This further analysis resulted in posteriors that peak at $\varepsilon \sim 1.7$ and $\Delta t \sim 0.015$~seconds, with $\ln\mathcal{B}_{g0}=7.5$, overwhelmingly favoring the glinted model over the model without a glint (see the left panel of Fig.~\ref{fig:GW190521-followup} in orange).
This result can be interpreted in two separate ways: either a glint somehow arrives before the main signal, or a glint arrives after the main signal but is somehow larger in amplitude than the main signal.
However one chooses to interpret this result, these recovered parameters do not comport with our theoretical model for how glints form.

We next explored an alternative waveform model, NRSur7dq4, which is a surrogate model for precessing binary black hole numerical relativity simulations \cite{NRSur7dq4}, to test if this result is due to a peculiarity with the template waveform family.
An analysis using NRSur7dq4 waveform templates also recovered similar posterior distributions in $\Delta t$ and $\varepsilon$, as shown in the left panel of Fig.~\ref{fig:GW190521-followup} in gray.
If this result is due to waveform systematics, then the systematics plague multiple waveform approximations.

Finally, we performed parameter estimation on the Livingston (L1) and Hanford (H1) detectors individually to test that the glint-parameter posteriors are consistent in each observing detector.
We found that an analysis using just L1 data (right panel of Fig.~\ref{fig:GW190521-followup} in orange) recovers a peak consistent with our two-detector result (gray); however, analyzing H1 data independently does not recover consistent parameters (blue).

Since the recovered glint features are not consistent with our theoretical model for glints and the feature is only found in L1, we therefore conclude that there is likely some noise artifact in the L1 detector that the glint templates are fitting, resulting in an increased evidence value relative to templates without glints.
Coincidentally, the period of the signal at merger is roughly 0.015~seconds, which is consistent with the recovered $\Delta t$ (left panel of Fig.~\ref{fig:GW190521-followup}).
This fact likely contributes to the noticeable increase in evidence when using templates with glints.

\clearpage

\bibliography{main.bib}
 
\end{document}